\documentstyle[12pt]{article}
\begin{document}

\begin{center}
\Large{\bf Volcanogenic Dark Matter 
           and Mass Extinctions } 
\end{center}
\vskip 3 cm 

\begin{center}
{\bf Samar Abbas $^{* \#}$ and Afsar Abbas $^\#$}

\vskip 5 mm

$\#$ Institute of Physics, Bhubaneswar-751005, India

$*$ Physics Dept., Utkal University, Bhubaneswar-751004, India

(e-mail : abbas@iopb.ernet.in, afsar@iopb.ernet.in)
\end{center}

\vskip 30 mm

\begin{centerline}
{\bf Abstract }
\end{centerline}
\vskip 5 mm

The passage of the Earth through dense clumps of dark matter, the
presence of which are predicted by certain cosmologies,
would produce large quantities of heat in the interior of 
this planet through the capture and subsequent annihilation of dark
matter particles. This heat can cause large-scale volcanism
which could in turn have caused the extinction of the dinosaurs
and other mass extinctions. The periodicity of such volcanic  
outbursts agrees with the frequency of palaeontological mass
extinctions as well as the observed periodicity in the occurrence of 
the largest flood basalt provinces on the globe.

\newpage
While investigating the possibility that a certain weakly
interacting massive particle (WIMP) , the cosmion, could explain
both the dark matter (DM) problem and the solar neutrino problem,
Press and Spergel 
  [ 1 ]    
estimated  the rate at which the Sun or a planet will capture WIMPs.
For the Earth the capture rate is , as given by Krauss et al
  [ 2 ]:

\begin{equation}
 \dot{N_{E}} = ( 4.7 \times 10^{17} sec^{-1} ) 
         \{ 3ab \}  
         \left[ \frac { \rho_{0.3} \sigma_{N,32} }
                      { \bar{v}_{300}^{3} } 
                \right] 
         ( \frac {1}{ 1 + m_{X}^{2} / m_{N}^{2} } ) \\
\end{equation}
 
\noindent where 
$ m_X $ is the mass of the DM particle, 
$ m_N $ is the mass of a typical nucleus off which the the particle
  elastically scatters with cross-section $ \sigma_N $,
$ \rho_X $ is the mean mass density of DM particles in the 
  Solar System,
$ \bar{v} $ is the r.m.s. velocity of dark matter in the Solar System, 
$ \rho_{0.3} =  \rho_X / 0.3 GeV cm^{-3}  $,
$ \sigma_{N,32} =  \sigma_N / 10^{-32} cm^2  $,
$ \bar{v}_{300} =  \bar{v} / 300 km s^{-1}  $, 
and
$ a $ and $ b $ are numerical factors of order unity which depend on the
density profile of the Sun or planet. 

The Earth will continue to accrete more and more particles until their
number density inside the planet becomes so high that they start to
annihilate. This could lead to a flux of neutrinos 
[ 2-4 ].

Depending on the nature of the dark matter ( neutralino, photino, 
gravitino, sneutrino, Majorana neutrino, etc. ), different 
annihilation channels are possible 
[ 3-5 ]. 
Generally the most significant channels are
$ \chi \bar{\chi} \rightarrow  q \bar{q} $ ( quark-antiquark ), 
$ \chi \bar{\chi} \rightarrow \gamma \gamma $ ( photons ) and
$ \chi \bar{\chi} \rightarrow l \bar{l} $ ( lepton-antilepton ).  
In the quark channel hadronization will take place through jets and
subsequent radiative decay will lead to mesons which in turn will decay 
through their available channels. Hence
[ 5 ] :
\begin{equation}
\chi \bar{\chi} \rightarrow q \bar{q} \rightarrow 
  ( \pi^{0}, \eta, ... ) 
  \rightarrow \gamma + Y  
\end{equation} 

\noindent All annihilation processes 
which directly or indirectly create
photons and where energy is delivered to the core through inelastic 
collisions would lead to the generation of heat in the core.
We wish to study this heat generation in the Earth's core
through annihilation. This heat is :
\begin{equation} 
 Q_E = e \dot{ N_E } m_X 
\end{equation} 

\noindent where 
e is the fraction of annihilations which lead to the generation of 
heat in the core of the Earth.
Note that e may be as large as unity for the ideal case where
the WIMPs annihilate predominantly through photons only. We
shall however take it to be $ \sim 0.5 $ [ 3-5 ] for an order
of magnitude estimate. 

Taking $ ab \sim 0.34 $ 
[2],
$ \rho_{0.3} =1 $, 
$ \bar{v}_{300} = 1 $, 
$ m_{X} = 55 GeV $
and the cross-section on iron to be $ \sigma_N = 10^{-33} cm^2 $,
we find that $ \sim 10^8 W $
of heat is generated. 

This estimate for the heat generation in the core of the earth is
for an isotropic DM distribution. However, clumpiness may be a
generic characteristic of CDM ( cold dark matter ) cosmologies. 
Recently it has been 
argued that in the cosmic string, texture and inflationary models
CDM clump cores are formed with core densities given by
$ \rho_c \sim 4.5 \times 10^{8} \rho_X \Omega^{2} \times h^4 $ 
[ 6,7 ]. 
These cores occur over a 
a smooth background halo density $\rho_X$. Although the core
densities can exceed the local halo density by a factor of $10^{10}$
[ 7 ], one may take the conservative value of 
$ \rho_c \sim 10^8 \rho_X $.
These clumps are expected to cross earth with a periodicity of 30-100
Myrs
[ 8 ]. 
Thus during the passage of the Earth through such clumps at 
regular intervals, the flux of the incident DM particles will
increase by roughly a factor of $\sim 10^8$. Consequently the value
of $ \dot{Q}_{E} $ during the passage 
of a clump will be $ \sim 10^{16} $ W.

Improving upon the previous work, Gould 
[ 9 ] 
obtained greatly enhanced capture rates for the Earth ( 10-300 times
that previously believed ) when the WIMP mass roughly equals the
nuclear mass of an element present in the Earth in large 
quantities, thereby constituting a resonant enhancement.
Gould's formula gives the capture rate for each element in the
Earth as 
[ 9 ] : 

\begin{equation}
\dot{ N_{E} } = ( 4.0 \times 10^{16} sec^{-1} )
         \bar\rho_{0.4}   
         \frac { \mu } { \mu_{+}^{2} }
         Q^{2}
         f
         \left< \hat\phi 
                ( 1- \frac { 1-e^{-A^2} } { A^2 } )
                \xi_{1} (A)
                \right>
\end{equation}

\noindent where
$ \bar\rho_{0.4} $ is the halo WIMP density normalized to 
   $ 0.4 GeVcm^{-3} $ ,
 Q = N - ( 1 - 4 $ sin^2 \theta_W $ ) Z
   $\sim$ N - 0.124Z,
 f is the fraction of the Earth's mass due to this element,   
$ A^2 = ( 3 v^2 \mu ) / ( 2 \hat{v}^2 \bar\mu_{-} ) $,
$ \mu =  m_{X} / m_{N}  $,
$ \mu_{+} = ( \mu + 1 ) / 2  $,
$ \mu_{-} = ( \mu - 1 ) / 2  $,
$ \xi_{1} (A) $ is a correction factor,
$ v = $ escape velocity at the shell of Earth material , 
$ \hat{v} = 3kT_w/m_X 
          = 300 kms^{-1} $ is the velocity dispersion, and 
$ \hat\phi =  v^2 / { v_{esc} }^2  $        
  is the dimensionless gravitational potential. 

  In the WIMP mass range 15 GeV-100 GeV this yields total
capture rates of the order of $ 10^{17} sec^{-1} $ to 
$ 10^{18} sec^{-1} $ .
According to the equation above, this yields
$Q_E \sim 10^8 W - 10^{10} W $ for a uniform density 
distribution. 

  In the case of clumped DM with core densities $ 10^8 $
times the galactic halo density, global power production due
to the passage of the Earth through a DM clump is
$ \sim 10^{16} W - 10^{18} W $. It is to be noted that this
heat generated in the core of the Earth is huge and arises
due to the highly clumped CDM.

Within the geothermodynamic theory it is believed that continuous 
heat absorption
by the D'' layer ( the lowermost layer of the mantle ) would
result from a temporary increase in heat transfer from the core
[ 10 ].
This process is thought to continue until, due to its 
decreasing density, this layer becomes
unstable, eventually breaking up into rising plumes
( plume production being an efficient way
of heat transfer ).  
This viewpoint of a lower mantle origin for plumes is 
strengthened by several recent observations. 
Firstly, high levels of He-3 reported for Siberian flood 
basalts 
[ 11 ] 
support this view, primordial He-3 having been retained within 
the lower mantle. Secondly, high levels of Osmium-187 
( a decay product of the element Rhenium-187 which is 
thought to
exist in high concentrations in the Earth's core ) 
observed in Siberian flood basalts 
[ 12 ] 
suggest that some of these rocks may even come from the
outer core.  

 A typical plume created in this manner would, due to its
lower density, well upwards. In this process, decompression of
the plume on account of its ascent in a pressure gradient will
lead to partial melting of the plume head, thereby producing 
copious amounts of basaltic magma
[ 13 ].
Mantle velocities being 
$ \sim 1 $ m/year , such a plume would take $ \sim 5 $ million
years to reach the crust. It would then melt its way through 
the continental crust, thereby producing viscous acidic
( silicic ) magma
[ 13 ].

The ultimate arrival of such a plume head at the surface could
be cataclysmic. Initial explosive silicic volcanism would be 
followed by periods of large-scale basalt volcanism that
ultimately lead to the formation of massive flood basalt
provinces such as the Siberian Traps, the Deccan Traps in India
and the Brazilian Paran\'{a} basalts.
Extensive atmospheric pollution would follow; the Deccan Trap
flood basalt volcanic episode ( $ \sim $ 65 million years ago )
ejected an estimated total of
$ 11 \times 10^{6} km^3 $ of basalt, 
$ 17 \times 10^{12} $ tonnes of $ H_2SO_4 $,  
$ 27 \times 10^{10} $ tonnes of $ HCl $, and
$ 19 \times 10^{3} km^3 $ of fine dust
[ 14 ]. 
Climatic models predict that this is capable of triggering
a chain of events ultimately leading to the depletion of the
ozone layer, global temperature changes, acid rain
and a decrease in surface ocean alkalinity.

Thus, Deccan volcanism has been proposed as a possible 
cause for the K/T ( Cretaceous/Tertiary ) mass extinction
that extinguished the dinosaurs   
[ 13,15 ], while
the Siberian basalts have been put forth as a 
possible culprit for the P/T ( Permian/Triassic )
mass extinction
[ 12 ]. In fact, there exists a striking concordance 
between the ages
of several major flood basalt provinces and the dates of
the major palaeontological mass extinctions
[ 10 ].

It was shown above that
the passage of the Earth through a DM clump core could produce
$ \sim 10^{16}-10^{18} W $ of heat. 
The annual heat flux of the Earth amounting to
$ 4.2 \times 10^{13} W $ [ 16 ], this would have
significant geothermodynamical effects which future work
shall shed more light on. Here, however, we suggest that this large
excess of heat produced in the interior of the Earth could lead
to violent volcanism and mass extinctions via the process elucidated
above.

In a recent paper, Collar set forth the hypothesis 
[ 8 ],
that dark matter induces cancers in organisms, thereby
causing mass extinctions. 
This hypothesis fails to
explain several features associated with the mass
extinction record. The iridium anomaly, the occurrence of soot, 
spherules, and other features at the K/T boundary 
generally attributed either to
an impact or a massive volcanic outburst
[ 17,18 ] find no explanation within this scenario.    
Moreover, sea level regression 
thought to occur at several mass
extinctions cannot fit in this model [ 8 ]. 
The occurence of several large flood basalt provinces with
ages concordant with several mass extinctions cannot be
visualized therein [ 8 ]. 
 
The volcanic hypothesis, despite providing a viable 
explanation for several features reported for mass 
extinctions,  
has always lacked a compelling 
reason for otherwise supposedly haphazard eruptions 
to occur in a periodic 
fashion. When one takes into account that the Earth has been cooling 
ever since its formation ( which implies a consequent decrease 
in volcanic activity ), this objection becomes a serious
weakness. 
It is
hoped that a viable reason for large volcanic eruptions
to occur in a periodic manner has been presented here
with the introduction of the volcanogenic dark matter scenario. This
should strengthen the volcanic
hypothesis of mass extinctions. 

\vskip 1 cm
Correspondence and requests for materials to either authors

Samar Abbas; e-mail : abbas@iopb.ernet.in

Afsar Abbas; e-mail : afsar@iopb.ernet.in  

\newpage
{\bf References}

1. Press W. H. and Spergel D. N., Ap. J. {\bf 296}, 679-684 (1985)

2. Krauss L. M., Srednicki M. and Wilczek F., Phys. Rev. {\bf D33},
2079-2083 (1986)

3. Gaisser T. K., Steigman G. and Tilav S., Phys. Rev. {\bf D34},
2206-2222 (1986)

4. Freese K., Phys. Lett. {\bf B167}, 295-300 (1986)

5. Bengtsson H-U., Salati P. and Silk J., Nucl. Phys. {\bf B346},
129-148 (1990)

6. Silk J. and Stebbins A., Ap. J., {\bf 411}, 439-449 (1993)

7. Kolb E. W. and Thachev I. I., Phys. Rev.{\bf D50}, 769-773 (1994)

8. Collar J. I., Phys. Lett. {\bf B368}, 266-269 (1996)

9. Gould A., Ap. J. {\bf 321}, 571-585 (1987)

10. Courtillot V. E., Sc. Am., 85-92 (Oct. 1990)

11. Basu A. R., Poreda R. J., Renne P. R., Teichmann F., Vasilev 
Y. R., Sobolev N. V. and Turrin B. D., Science {\bf 269}, 822-825
(1995)

12. Walker R. J., Morgan J.W., Horan M.F., Science {\bf 269},
819-822 (1995)

13. Campbell I. H., Czamanske G. K., Fedorenko V. A., Hill R. I.
and Stepanov V., Science {\bf 258}, 1760-1763 (1992)

14. Officer C. B.,Hallam A., Drake C. L. and Devine J. D., Nature
{\bf 326}, 143-149 (1987)

15. Officer C. and Page J., " The Great Dinosaur Controversy ",
Addison-Wesley (1996)

16. Vacquier V., Geophys. J. Int. {\bf 106}, 199-202 (1991) 

17. Bhandari N., Current Sc. (India) {\bf 61}, 97-104 (1991)

18. Sutherland F. L., Earth-Sc. Rev. {\bf 36}, 1-26 (1994)

\end{document}